\documentclass[11pt, a4paper, twoside]{fec}
\usepackage{etoolbox}
\patchcmd\thebibliography
{\labelsep}
{\labelsep\itemsep=-4pt\relax}
{}
{\typeout{Couldn't patch the command}}

\title{Effect of magnetic perturbations on the 3D MHD self-organization of shaped tokamak plasmas}

\author[1]{D. Bonfiglio}
\author[1]{S. Cappello}
\author[1]{M. Veranda} 
\author[2]{L. Chac\'on}
\author[3,1]{D. F. Escande}
\affil[1]{Consorzio RFX, 35127 Padova, Italy}
\affil[2]{Los Alamos National Laboratory, Los Alamos, New Mexico 87545, USA}
\affil[3]{Aix-Marseille Universit\'e, CNRS, PIIM, UMR 7345, 13013 Marseille, France}

\doi

\emailID{daniele.bonfiglio@igi.cnr.it}

\begin{document}

\maketitle

\begin{abstract}
The effect of magnetic perturbations (MPs) on the helical self-organization of shaped tokamak plasmas is discussed in the framework of the nonlinear 3D MHD model. Numerical simulations performed in toroidal geometry with the \textsc{pixie3d} code [L. Chac\'on, Phys. Plasmas {\bf 15}, 056103 (2008)] show that $n=1$ MPs significantly affect the spontaneous quasi-periodic sawtoothing activity of such plasmas. In particular, the mitigation of sawtooth oscillations is induced by $m/n=1/1$ and $2/1$ MPs. These numerical findings provide a confirmation of previous circular tokamak simulations, and are in agreement with tokamak experiments in the RFX-mod and DIII-D devices. Sawtooth mitigation via MPs has also been observed in reversed-field pinch simulations and experiments. The effect of MPs on the stochastization of the edge magnetic field is also discussed.
\end{abstract}

\section{Introduction}


Three-dimensional (3D) effects play an important role in the physics of nominally axisymmetric toroidal configurations for the magnetic confinement of fusion plasmas, such as the tokamak and the reversed-field pinch (RFP). 3D effects can result from plasma self-organization, in particular when a magnetohydrodynamic (MHD) instability nonlinearly saturates and gives rise to a long-lived helical state. This occurs in the RFP with the spontaneous emergence of the single-helical-axis (SHAx) state with improved confinement properties \cite{lorenzini2009np}, and in the tokamak with the so-called density snake \cite{weller1987prl} and the long-lived helical states in advanced regimes \cite{graves2013ppcf}. On the other hand, 3D effects can be induced by external magnetic perturbations (MPs) produced with non-axisymmetric coils. Notable tokamak applications of non-axisymmetric coil systems, traditionally used for error field correction and feedback control of MHD instabilities \cite{chu2010ppcf}, are edge-localized modes (ELMs) suppression \cite{evans2004prl}, drive of neoclassical toroidal rotation \cite{garofalo2008prl}, and Alfv\'en modes mitigation \cite{bortolon2013prl}. In addition, spontaneous and externally applied 3D effects can couple. In fact, it is possible to stimulate and control the spontaneous helical self-organization with the application of external 3D fields. This has been first demonstrated in the RFP configuration. In the RFX-mod experiment \cite{sonato2003fed}, the persistence of spontaneous SHAx states is increased by applying MPs with the same helicity as the core helical structure \cite{piovesan2011ppcf}. It is also possible to stimulate helical states under conditions where they do not form spontaneously, such as at low current or high density \cite{piovesan2014nf}. Moreover, by applying MPs with helicities different from the spontaneous one, helical states with the chosen helicity can be induced \cite{cappello2012iaea,martin2013nf}. In all the above cases, MPs are found to mitigate the spontaneous sawtooth activity of RFP plasmas \cite{piovesan2013pop}. This observation has triggered similar tokamak experiments in RFX-mod operated as a circular Ohmic tokamak and then in DIII-D. In both devices, MPs with $n=1$ toroidal periodicity have been shown to mitigate the sawtooth activity associated with the internal kink mode \cite{bonfiglio2013eps,martin2014iaea}. This evidence points to the common physics basis of RFP and tokamak plasmas.


The nonlinear visco-resistive 3D MHD model is successful in qualitatively reproducing the above-mentioned experimental findings. In the RFP case, the experimental discovery of persistent helical states \cite{escande2000ppcf} has been stimulated by the spontaneous helical self-organization occurring in nonlinear 3D MHD simulations with ideal boundary conditions when the visco-resistive dissipation is increased \cite{cappello1992pof,cappello2000prl,cappello2004ppcf}. A closer agreement with respect to experimental observations has been obtained after the modification of the boundary conditions of numerical tools to allow for MPs of the plasma boundary \cite{bonfiglio2011nf}. With the application of MPs, the parameter space of helical RFP solutions is substantially enlarged towards small visco-resistive dissipation conditions. Numerical RFP studies with MPs have reproduced both the increased sawtoothing frequency and spontaneous helical state persistence \cite{veranda2013ppcf,bonfiglio2013prl,bonfiglio2015ppcf} and the stimulation of helical states with different helicities \cite{veranda2013ppcf}. Qualitative agreement with respect to experimental findings is obtained in tokamak modelling as well, where sawtooth mitigation and helical self-organization triggered by MPs are observed \cite{bonfiglio2013eps,bonfiglio2015ppcf}. In this report, we extend previous tokamak studies performed by assuming a circular axisymmetric equilibrium, by investigating the effect of MPs in D-shaped diverted tokamak configurations. This is motivated on the one hand by a closer comparison with sawtooth mitigation experiments in DIII-D, and on the other hand by a more general effort of numerical tools development towards realistic conditions.


The paper is organized as follows. After a description of the employed numerical model in Sec. \ref{sec:model}, numerical results are reported in Sec. \ref{sec:results}. Two shaped axisymmetric equilibria are described in Sec. \ref{sec:axisymmetric_equilibria}, one elongated and one D-shaped configuration. In Sec. \ref{sec:axisymmetric_simulations}, axisymmetric MHD simulations based on these equilibria are discussed. In particular, it is shown that the purely elongated configuration is unstable with respect to vertical displacements of the plasma, while the D-shaped configuration is vertically stable. 3D MHD simulations starting from the D-shaped equilibrium are considered in the rest of the paper. In Sec. \ref{sec:3d_simulations_wo_helical_mps}, simulations without helical MPs are reported. It is shown that with relatively small dissipation parameters a sawtoothing dynamics is obtained, while the bifurcation to a stationary equilibrium with helical core occurs at larger dissipation. In Sec. \ref{sec:3d_simulations_with_helical_mps}, the application of helical MPs in the case with spontaneous sawtooth oscillations is considered. $n=1$ MPs with varying poloidal periodicity $m$ are applied, and the effect on the sawtoothing dynamics is analysed. Final remarks and future perspectives are discussed in Sec. \ref{sec:conclusion}.

\section{Numerical model}
\label{sec:model}


The numerical simulations reported in this paper are performed in toroidal geometry with the nonlinear 3D MHD code \textsc{pixie3d}, which solves the compressible visco-resistive MHD equations with finite $\beta$ and heat transport \cite{chacon2004cpc,chacon2008pop}. In this report, we run the code in the constant-density and zero-$\beta$ approximation, whose equations is dimensionless form are given by
\begin{equation}
\label{eq:mhd_a}
\frac{\partial{\bf B}}{\partial t}=-\nabla\times{\bf E},\quad {\bf E}=\eta{\bf J}-{\bf v}\times{\bf B},\quad {\bf J}=\nabla\times{\bf B},\quad \nabla\cdot{\bf B}=0,
\end{equation}
\begin{equation}
\label{eq:mhd_b}
\rho\left[\frac{\partial{\bf v}}{\partial t}+({\bf v}\cdot\nabla){\bf v}\right]={\bf J}\times{\bf B} + \rho\nu\nabla^2{\bf v}.
\end{equation}
Here, $t$ is the time, $\rho$ the uniform particle density, ${\bf v}$ the plasma velocity, ${\bf B}$ the magnetic field, ${\bf E}$ the electric field, ${\bf J}$ the current density, $\eta$ the resistivity, and $\nu$ the kinematic viscosity. The magnetic field is normalized to the on-axis toroidal magnetic field $B_0$, the density to the on-axis ion mass density $\rho_0$, lengths to the torus minor radius $a$, velocity and time to the on-axis Alfv\'en speed $v_\mathrm{A}=B_0/\sqrt{\mu_0 \rho_0}$ and Alfv\'en time $t_\mathrm{A}=a/v_\mathrm{A}$. In these units, the on-axis resistivity and viscosity correspond to the inverse Lundquist number $\eta_0=\tau_A/\tau_R \equiv S^{-1}$ and to the inverse viscous Lundquist number $\nu_0 = \tau_A/\tau_V \equiv M^{-1}$, $\tau_R$ and $\tau_V$ being the resistive and viscous time scales. In this report, resistivity and viscosity are assigned as constant parameters. In particular, $\nu$ is assumed to be uniform, while $\eta$ is a prescribed function of the radius, to be discussed later. No momentum source is included in Eq. \ref{eq:mhd_b} and all the perturbed harmonics have the same angular phase, so that the net torque between MHD modes and the resulting mean plasma flow are zero. Eqs. \ref{eq:mhd_a}-\ref{eq:mhd_b} are actually the same equations solved by the \textsc{specyl} code in cylindrical geometry \cite{cappello1996nf,cappello2004ppcf}. The nonlinear verification benchmark between \textsc{specyl} and \textsc{pixie3d} in their common limit of application has demonstrated that the two codes provide a highly accurate and reliable solution of the 3D nonlinear visco-resistive MHD model \cite{bonfiglio2010pop}.


\textsc{pixie3d} uses a finite-volume scheme for the spatial discretization of MHD equations and fully-implicit temporal discretization schemes such as second-order Crank-Nicolson and backward-differentiation-formula (BDF2) \cite{chacon2004cpc,chacon2008pop}. A preconditioned Newton-Krylov method is employed to solve the resulting set of nonlinear algebraic equations. The parallel version of the code features excellent scalability properties \cite{chacon2008pop}. \textsc{pixie3d} is written in a generalized curvilinear formulation \cite{chacon2004cpc}, which makes the code applicable to different geometries, in particular cylindrical and toroidal of interest for fusion plasmas simulations. The numerical simulations reported in this paper are performed in toroidal geometry with circular cross-section and coordinates $(r,\theta,\phi)$, where $r$ is the radius and $\theta$ and $\phi$ the poloidal and toroidal angles. $R_0$ and $a$ are the major and minor radii of the torus, so that $R=R_0 + a \cos \theta$ and $Z=a\sin\theta$ are the positions with respect to the central axis and the equatorial plane. We are going to study shaped tokamak plasmas with aspect ratio $A\simeq 2.5$, relevant to present-days diverted tokamaks such as DIII-D, ASDEX-Upgrade and JT60-SA. To get this plasma aspect ratio, the aspect ratio of the circular toroidal domain is set to $R_0/a=1.65$. 


The imposed boundary conditions at $r=a$ are such that $E_\theta(a)=0$, which implies that the total toroidal magnetic flux $\Phi_0$ is conserved in time. The edge toroidal electric field is $E_\phi(a)=E_0/R$, where $E_0$ is the constant externally imposed toroidal electric field. No-slip boundary conditions are used for the edge tangential velocity, while the edge radial velocity is $v_r(a)=E_0({\bf e}_\phi\times{\bf B})\cdot{\bf e}_r/B^2$ \cite{delzanno2008pop}. From the radial component of the Faraday equation $\frac{\partial{\bf B}}{\partial t}=-\nabla\times{\bf E}$ together with the boundary conditions on $E_\theta$ and $E_\phi$, it follows that the edge radial magnetic field $B_r(a)$ is constant in time. The desired boundary condition on $B_r(a)$ is obtained with a proper initial perturbation of the total magnetic field, which must satisfy $\nabla\cdot{\bf B}=0$ and whose edge radial component will be conserved in time. The initial magnetic perturbation is prescribed in terms of the toroidal periodicity $m$, poloidal periodicity $n$ and amplitude of the corresponding Fourier harmonics in the geometrical angles $\theta$ and $\phi$. In this paper, static $B_r(a)$ perturbations are applied with both $n=0$ and $n\ne 0$ toroidal periodicities. In particular, large $n=0$ MPs provide the target axisymmetric plasma shaping, while smaller $n\ne 0$ perturbations mimic the effect of non-axisymmetric coils. We note that, from the physical point of view, the above boundary conditions on the edge electric and magnetic fields can be regarded as a continuous distribution of saddle coils located just inside a perfectly conducting wall. 


The initial condition of \textsc{pixie3d} simulations is provided by the iterative solution of the Grad-Shafranov equation ${\bf J}\times{\bf B}=0$ and the parallel Ohm's law ${\bf E}\cdot{\bf B}=\eta {\bf J}\cdot{\bf B}$, with the assumption of a circular magnetic boundary $B_r(a)=0$. The externally imposed toroidal electric field $E_0$ and the radial resistivity profile are chosen in such a way that the initial safety factor profile takes a tokamak form, as also used for instance in Refs. \cite{furth1973pof,shan1993ppcf},
\begin{equation}
  \label{eq:qprofile}
  q(r)=q_0 \left\{1+\left[\left(\frac{q_a}{q_0}\right)^\lambda-1\right]r^{2\lambda}\right\}^\frac{1}{\lambda},
\end{equation}
where $q_0=\frac{2a\eta_0}{R_0 E_0}$ and $q_a$ are the values of the safety factor at $r=0$ and $r=a$, respectively, and the parameter $\lambda$ defines whether the resulting current profile is peaked or flattened. Throughout this paper, we set $q_0=0.8$ and $q_a=10$. The $\lambda$ parameter is assumed to be $\lambda=1$, corresponding to a peaked current profile.

\section{Shaped tokamak 3D MHD simulations}
\label{sec:results}

\subsection{Shaped axisymmetric equilibria}
\label{sec:axisymmetric_equilibria}


In this paper, tokamak simulations are reported based on two different shaped axisymmetric equilibria, obtained by initial $n=0$ magnetic field perturbations whose edge radial component will be conserved in time due to the above-mentioned boundary conditions. The first is a purely elongated equilibrium, with perturbation of the $m/n=2/0$ Fourier harmonic of the magnetic field with edge amplitude $\Im b_r^{2/0}(a)=8\%$ of the on-axis toroidal field $B_0$. The second is a D-shaped equilibrium which includes triangularity as well, and is obtained by the perturbation of both the $2/0$ harmonic with amplitude $\Im b_r^{2/0}(a)=4\%$ and the $3/0$ harmonic with $\Im b_r^{3/0}(a)=-8\%$. Only the imaginary parts of the $B_r$ harmonics are perturbed because in real space they correspond to the $\sin(m\theta)$ component, which is the one required to get up-down symmetric magnetic configurations.

\begin{figure}[t]
  \begin{center}
    \includegraphics[scale=1.25]{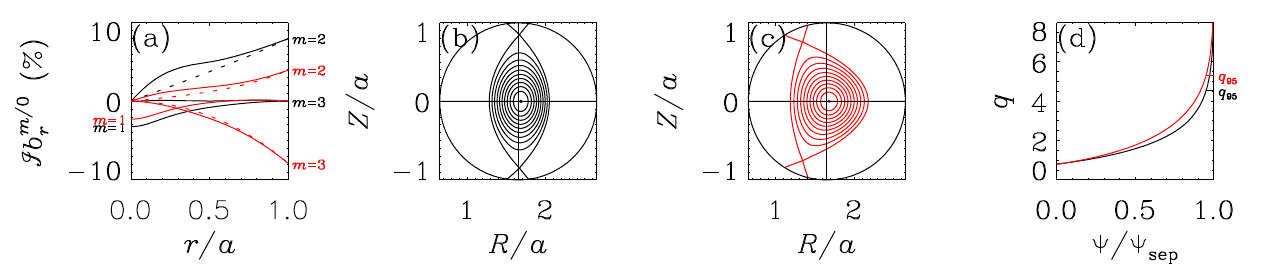}
    \caption{Shaped axisymmetric equilibria obtained at $t=100\ \tau_\mathrm{A}$ of \textsc{pixie3d} simulations with $S=10^6$ and $M=3.3\times 10^4$. (a) radial profiles of the imaginary part of the main $n=0$ harmonics of $B_r$, normalized to the on-axis toroidal field $B_0$, for the elongated case (black curves) and the D-shaped case (red). The dashed curves come from ``vacuum'' \textsc{pixie3d} simulations with $E_0=0$, $S=1$ and $M=1$. Contour levels of the poloidal flux $\Psi$ for the elongated case (b) and D-shaped case (c). (d) safety factor profile $q=\frac{\mathrm{d}\Phi}{\mathrm{d}\Psi}$ from the magnetic axis $\Psi=0$ to the separatrix $\Psi=\Psi_\mathrm{sep}$ for the elongated case (black curve) and the D-shaped case (red).}
    \label{fig:1}
  \end{center}
\end{figure}

To compute the resulting axisymmetric equilibria, \textsc{pixie3d} simulations with the two different initial axisymmetric perturbations are performed. The simulations use a 2D mesh with 128 points in the radial direction and 64 points in the poloidal one, and dissipation parameters $S=10^6$ and $M=3.3\times 10^4$. After around $100$ Alfv\'en times, a self-consistent equilibrium configuration is obtained in both cases. Such equilibrium configurations are shown in Fig. \ref{fig:1}. In both cases, the $1/0$ $B_r$ harmonic (representing the Shafranov shift) is peaked in the core, while the perturbed $2/0$ and $3/0$ harmonics increase towards the edge. The perturbed harmonics are compared with the corresponding vacuum profiles, obtained from \textsc{pixie3d} simulations with the same $B_r(a)$ perturbations but $E_0=0$, $S=1$ and $M=1$, resulting in negligible $\bf{J}$ and $\bf{v}$ fields. A significant plasma amplification with respect to the vacuum profile is observed for the $2/0$ harmonic. The amplitudes of the applied MPs are chosen so that a magnetic separatrix with two X-points (i.e., a double-null diverted configuration) is obtained in both cases. This is observed by looking at the shape of the magnetic surfaces defined by ${\bf B}\cdot\nabla\Psi=0$, where $\Psi=2\pi R A_\phi$ is the poloidal magnetic flux function and ${\bf A}$ is the vector potential in the $A_r=0$ gauge. In both cases, the O-point (i.e., the plasma magnetic axis) is outward shifted at $r/a\simeq0.05$ and the X-points are located around $r/a=0.8$. The volume inside the magnetic separatrix is larger in the D-shaped configuration. The corresponding safety factor profile $q(\Psi)=\frac{\mathrm{d}\Phi}{\mathrm{d}\Psi}$ is also shown, where $\Phi(\Psi)$ is the toroidal magnetic flux within the magnetic surface labeled by $\Psi$. The safety factor at the magnetic axis $\Psi=0$ is quite close to $0.8$, the on-axis value chosen for the circular $q(r)$ profile in the absence of axisymmetric MPs. On the contrary, at the edge the effect of the axisymmetric MPs is significant. In particular, the divergence of $q$ is observed at the magnetic separatrix $\Psi_\textrm{sep}$ where $\mathrm{d}\Psi=0$. The safety factor at 95\% of the normalized poloidal flux ($q_{95}$), is in the range between $4$ and $6$ as in typical experimental tokamak profiles.

\begin{figure}[t]
  \begin{center}
    \includegraphics[scale=1.25]{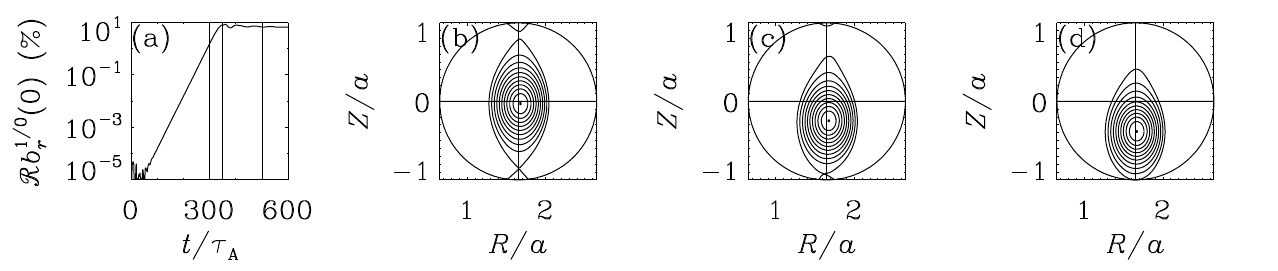}
    \caption{Elongated axisymmetric simulation with $S=10^6$ and $M=3.3\times 10^4$. (a) temporal evolution of the real part of the $1/0$ $B_r$ harmonic at $r=0$, normalized to the on-axis toroidal field $B_0$. (b-d) Contour levels of $\Psi$ at the three simulation times marked in panel (a).}
    \label{fig:2}
  \end{center}
\end{figure}

\begin{figure}[t]
  \begin{center}
    \includegraphics[scale=1.25]{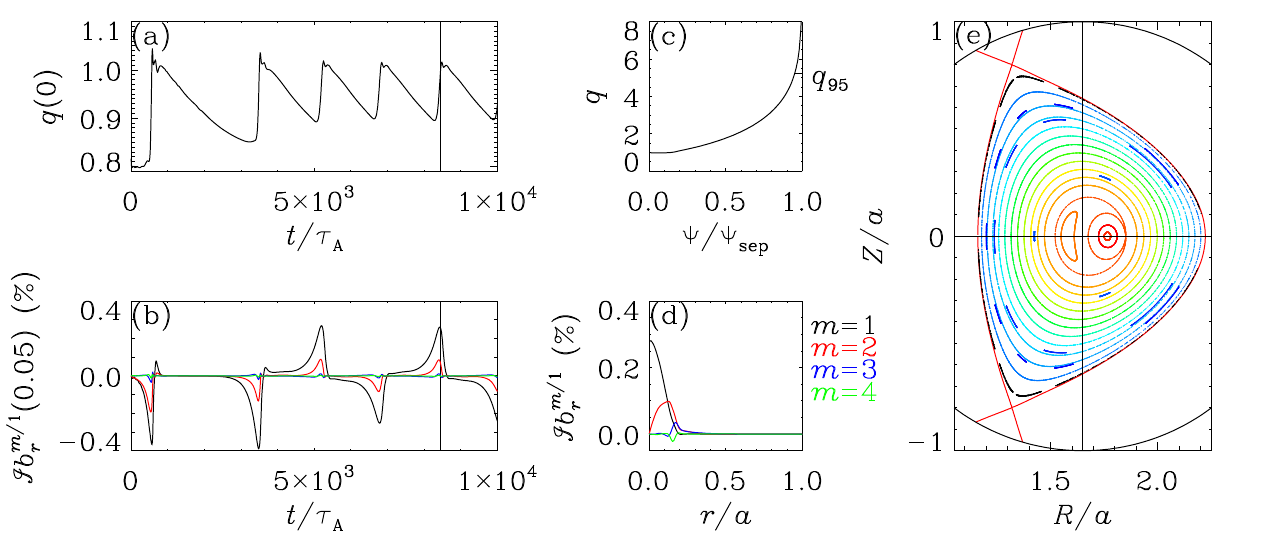}
    \caption{D-shaped 3D simulation with $S=10^6$ and $M=3.3\times 10^4$ and without helical MPs. (a) temporal evolution of the safety factor at the magnetic axis $\Psi=0$. (b) temporal evolution of the the imaginary part of the main $n=1$ harmonics of $B_r$ at $r/a=0.05$, normalized to the on-axis toroidal field $B_0$. For the simulation time marked in panels (a) and (b), it is shown (c) the safety factor profile (d) the radial profiles of the imaginary part of the $n=1$ $B_r$ harmonics and (e) the Poincar\'e plot at constant toroidal angle (colored dots) on top of the separatrix of the underlying axisymmetric configuration (red curve).}
    \label{fig:3}
  \end{center}
\end{figure}

\begin{figure}[t]
  \begin{center}
    \includegraphics[scale=1.25]{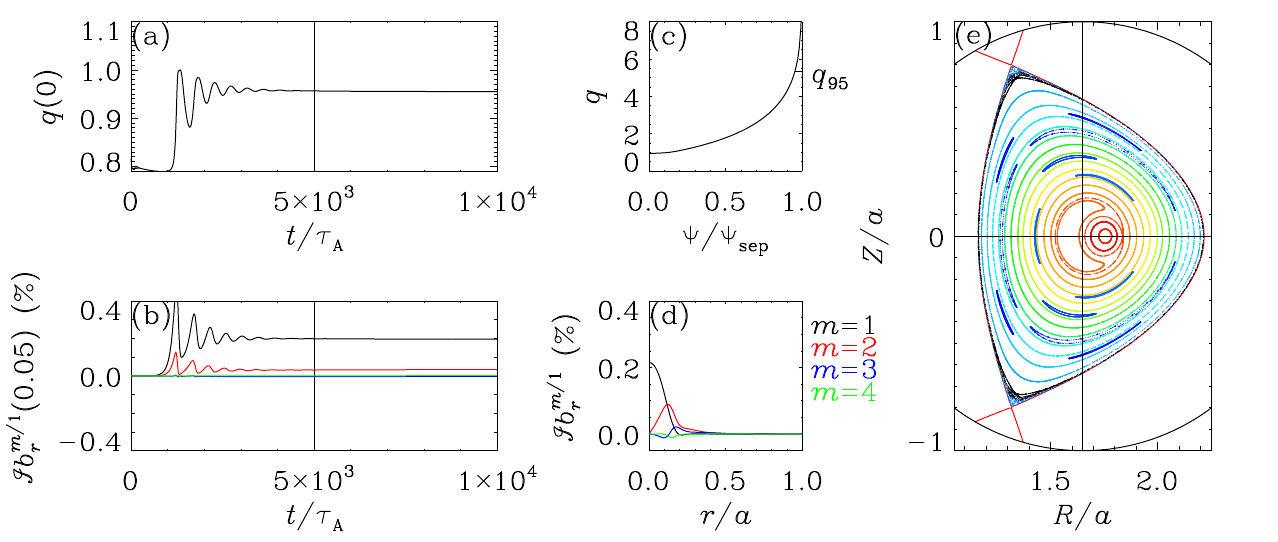}
    \caption{D-shaped 3D simulation with $S=10^5$ and $M=10^4$ and without helical MPs. Same quantities as in Fig. \ref{fig:3}.}
    \label{fig:4}
  \end{center}
\end{figure}

\begin{figure}[t]
  \begin{center}
    \includegraphics[scale=1.25]{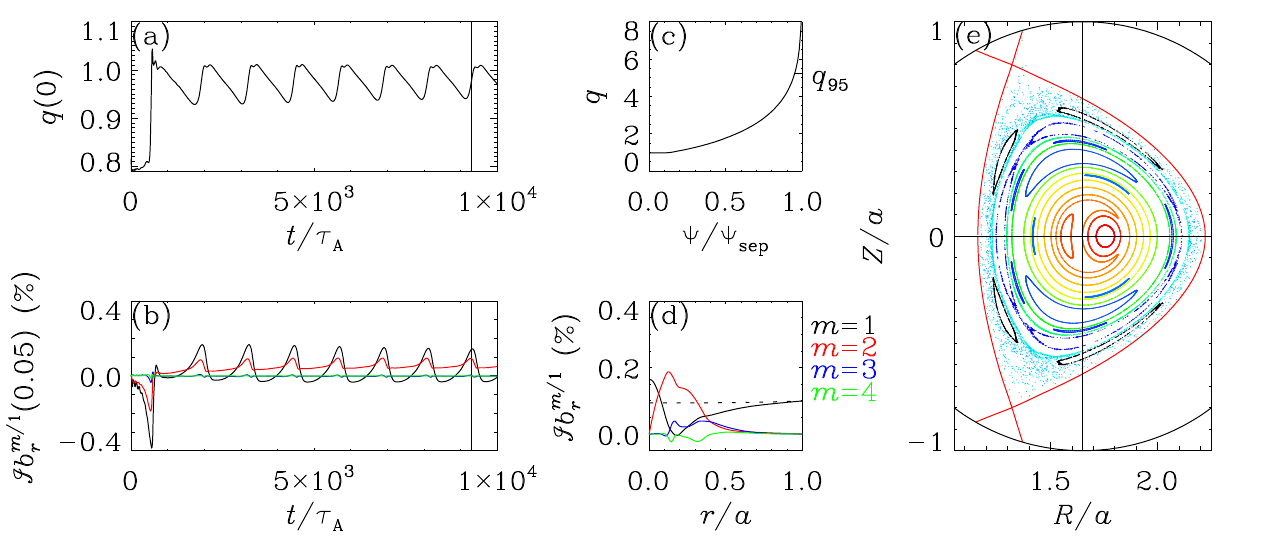}
    \caption{D-shaped 3D simulation with $S=10^6$, $M=3.3\times 10^4$ and $1/1$ MPs. Same quantities as in Fig. \ref{fig:3}. The dashed curve in panel (d) is the radial profile of the perturbed $B_r$ harmonic resulting from the corresponding ``vacuum'' \textsc{pixie3d} simulations with $E_0=0$, $S=1$ and $M=1$.}
    \label{fig:5}
  \end{center}
\end{figure}

\begin{figure}[t]
  \begin{center}
    \includegraphics[scale=1.25]{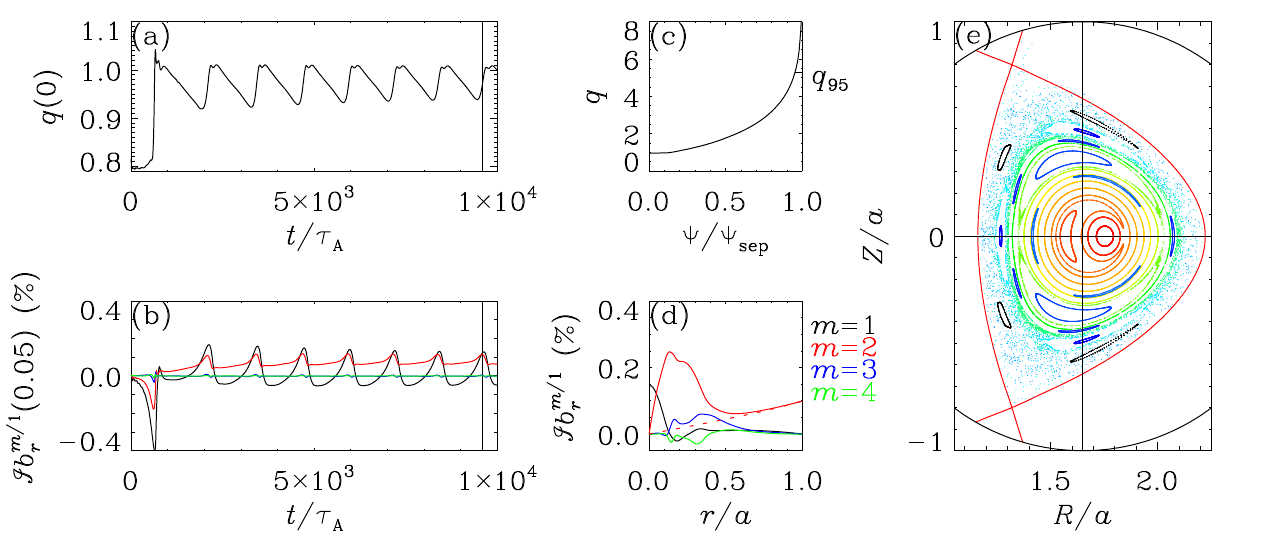}
    \caption{D-shaped 3D simulation with $S=10^6$, $M=3.3\times 10^4$ and $2/1$ MPs. Same quantities as in Fig. \ref{fig:5}.}
    \label{fig:6}
  \end{center}
\end{figure}

\begin{figure}[t]
  \begin{center}
    \includegraphics[scale=1.25]{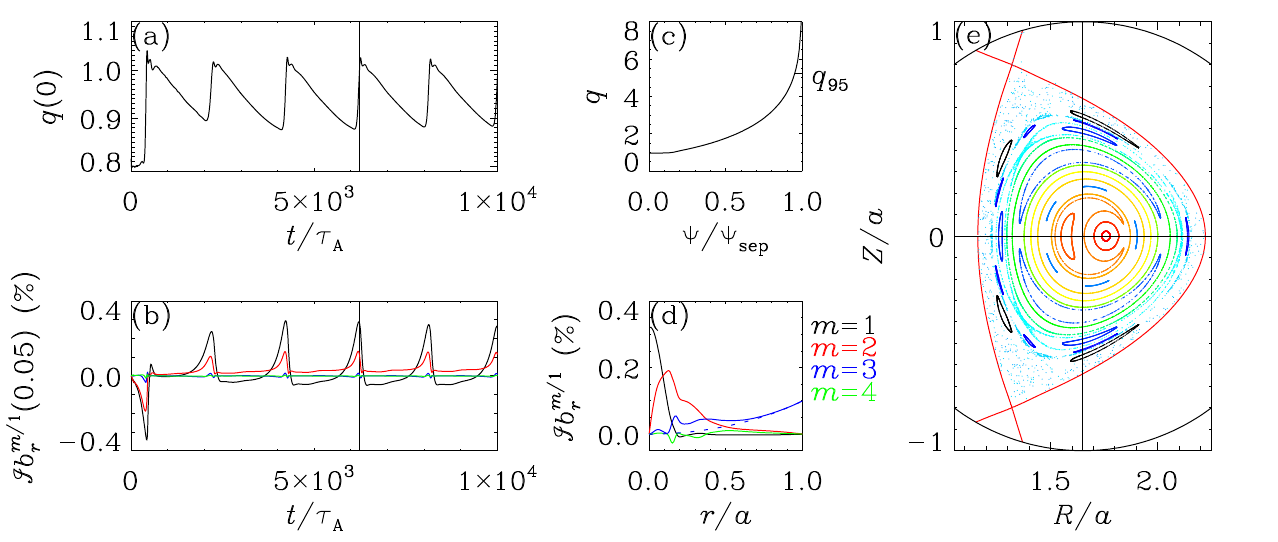}
    \caption{D-shaped 3D simulation with $S=10^6$, $M=3.3\times 10^4$ and $3/1$ MPs. Same quantities as in Fig. \ref{fig:5}.}
    \label{fig:7}
  \end{center}
\end{figure}

\begin{figure}[t]
  \begin{center}
    \includegraphics[scale=1.25]{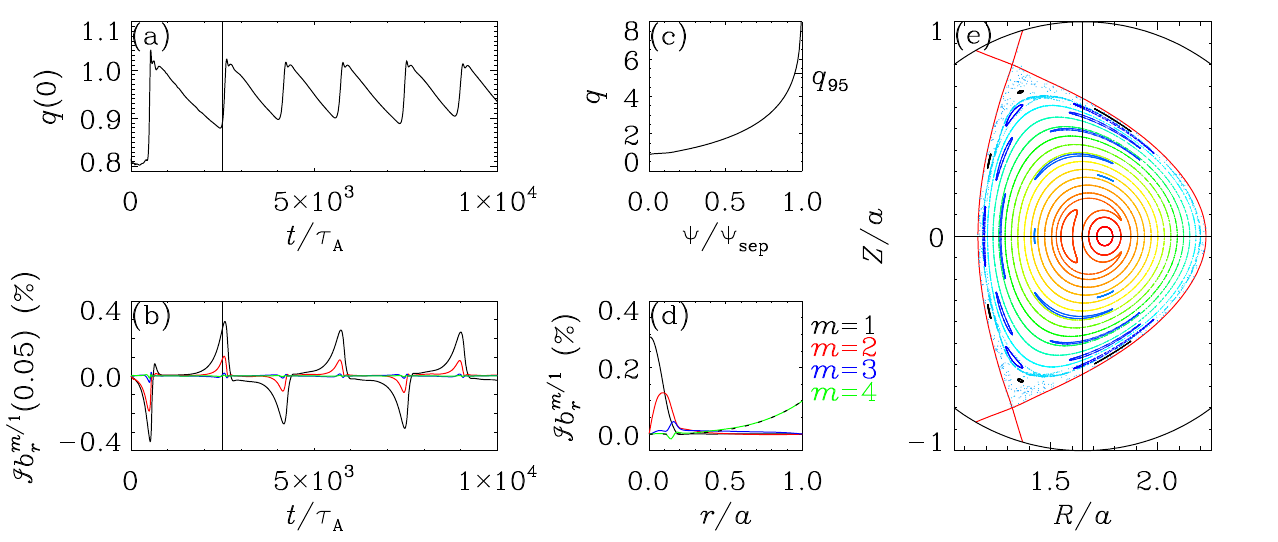}
    \caption{D-shaped 3D simulation with $S=10^6$, $M=3.3\times 10^4$ and $4/1$ MPs. Same quantities as in Fig. \ref{fig:5}.}
    \label{fig:8}
  \end{center}
\end{figure}

\subsection{Shaped axisymmetric simulations}
\label{sec:axisymmetric_simulations}


The purely elongated configuration turns out to be unstable with respect to the vertical instability, whose features are discussed in detail in Ref. \cite{wesson1978nf}. The corresponding \textsc{pixie3d} simulation with $2/0$ MP is reported in Fig. \ref{fig:2}. The real part of the $1/0$ $B_r$ harmonic is negligible at the beginning of the simulation, meaning that there is no $\cos(m\theta)$ component of $B_r$. However, such component grows exponentially on an inertial time scale. After a few hundreds Alfv\'en times, the linear phase ends and the instability is nonlinearly saturated. When looking at the shape of the magnetic surfaces, it is observed that around the end of the linear phase there is already a significant vertical displacement of the plasma but the lower X-point is still far from the wall. At the beginning of the saturated phase the lower X-point is rather close to the wall, which is eventually touched by the plasma with consequent loss of the diverted configuration. No vertical instability is observed in the case of the D-shaped configuration. In fact, the \textsc{pixie3d} simulation with both $2/0$ and $3/0$ MPs (whose temporal evolution is not shown here) predicts a stable $1/0$ vertical mode.


The properties of the two considered axisymmetric configurations are in qualitative agreement with the stability analysis for an elongated cylindrical plasma surrounded by an ideal wall as reported in Ref. \cite{wesson1978nf}. Such analysis predicts that the axisymmetric mode is stable provided that the elongation is sufficiently small and the plasma is close enough to the ideal wall. The elongated configuration considered in this report turns out to be marginally unstable according to the stability condition given in terms of the vertical and horizontal plasma radii and the distance from the wall. The fact that the D-shaped configuration is stable is also reasonable, since the elongation is lower in that case and, in addition, the effect of triangularity is stabilizing.

\subsection{D-shaped 3D simulations without helical MPs}
\label{sec:3d_simulations_wo_helical_mps}


We now discuss fully 3D simulations based on the D-shaped equilibrium configuration. These simulations use a 3D mesh with 128 points in the radial direction, 64 points in the poloidal angle and 64 points in the toroidal one. As discussed in Ref. \cite{bonfiglio2010pop}, a mesh with 64 angular points is adequate to support both $m=1$ and higher-$m$ harmonics. In this Section, simulations with no further MPs in addition to the $n=0$ required for axisymmetric plasma shaping are considered. We first describe a simulation with $S=10^6$ and $M=3.3\times 10^4$, as reported in Fig. \ref{fig:3}. Quasi-periodic sawtooth oscillations associated with the internal kink mode are observed in this case. The dynamics is similar to the one previously observed in circular tokamak simulations with \textsc{specyl} and \textsc{pixie3d} codes \cite{bonfiglio2010pop,bonfiglio2013eps,bonfiglio2015ppcf}. The safety factor at the magnetic axis $\Psi=0$ undergoes sawtooth oscillations in the range from $q(0)\simeq 0.9$ to $q(0)\simeq 1$. These oscillations are ruled by the nonlinear MHD dynamics of the internal kink mode, whose principal spectral component is the $1/1$ Fourier harmonic. The core amplitude of the $1/1$ $B_r$ harmonic grows during slow phases with decreasing $q(0)<1$ until a sudden jump to $q(0)\simeq 1$ is triggered. Then, the amplitude of the internal kink crashes and the cycle starts again. Only the imaginary part of the $B_r$ harmonics is shown since the real part is zero throughout the simulation. This is because the initial velocity perturbation that triggers the internal kink instability is applied in such a way to produce a $\sin(m\theta-n\phi)$ component in $B_r$ and, due to the properties of the visco-resistive MHD equations, this phase condition is conserved in time. The configuration at a time of maximum internal kink mode amplitude is characterized by a rather flat $q(\Psi)$ profile inside the $q=1$ rational surface. The radial profile of the $1/1$ $B_r$ harmonic is peaked in the core and goes to zero outside $r/a\simeq 0.2$. Other $n=1$ Fourier harmonics are also present, which in this reference simulation case are mainly due to the toroidal and geometric couplings with the principal $1/1$ harmonic. The corresponding Poincar\'e plot at constant toroidal angle, computed with the field-line tracing code \textsc{nemato} \cite{finn2005pop}, is characterized by the presence of a large magnetic island formed by the internal kink mode. The island is close to completely reconnecting the magnetic flux within the $q=1$ surface, after which nested magnetic surfaces would be recovered. Due to the phase of the seed magnetic fluctuation that would survive the crash, the island would grow on the opposite side of the magnetic axis during the next sawtooth cycle. This $\pi$ phase flip spontaneously occurs after each crash and is consistent with the sign reversal in the temporal evolution of the imaginary component of the core $1/1$ $B_r$ harmonic. Secondary islands with almost negligible width are observed in the Poincar\'e plot at major rational surfaces. They are transiently induced by the internal kink mode at its largest amplitude via the toroidal and geometric couplings.


As already shown by \textsc{specyl} and \textsc{pixie3d} tokamak simulations in cylindrical geometry, if the visco-resistive dissipation is increased a stationary configuration is obtained with a saturated island in the core \cite{bonfiglio2010pop}. The same feature is recovered here, as reported in Fig. \ref{fig:4} corresponding to a simulation with $S=10^5$ and $M=10^4$. In this case, the initial oscillations in $q(0)$ and in the core amplitude of the $1/1$ $B_r$ harmonic are damped and a stationary configuration with $q(0)\simeq 0.95$ and saturated internal kink mode is eventually obtained. The final equilibrium configuration is characterized by the presence of a partially reconnected magnetic island associated with the saturated internal kink mode. In this D-shaped case, geometry-induced secondary magnetic islands of small width are observed. This kind of stationary helical configuration may be relevant to experimental conditions where long-lived helical states are observed, such as the density snake \cite{weller1987prl} or saturated ideal modes in advanced tokamak regimes \cite{graves2013ppcf}. However, a helical core without a magnetic island is typically observed in such experimental conditions. As shown by finite-$\beta$ \textsc{pixie3d} simulations discussed in Ref. \cite{bonfiglio2015ppcf}, these experimental regimes may be better described by considering the pressure-driven branch of the internal kink mode.

\subsection{D-shaped 3D simulations with helical MPs}
\label{sec:3d_simulations_with_helical_mps}


We now focus on the simulation regime with small dissipation parameters and quasi-periodic sawtooth oscillations, and discuss the effect of applied MPs. In this report, MPs with $n=1$ toroidal periodicity and different poloidal periodicity from $m=1$ to $m=4$ are considered. This choice is consistent with previous circular tokamak studies \cite{bonfiglio2013eps,bonfiglio2015ppcf}. We first look at the case with a $1/1$ MP with amplitude $\Im b_r^{1/1}=0.1\%$ of the on-axis toroidal field $B_0$, as reported in Fig. \ref{fig:5}. Similarly to what shown for the circular tokamak \cite{bonfiglio2013eps,bonfiglio2015ppcf}, MPs with same periodicity as the internal kink mode mitigate the sawtoothing activity. In fact, the average period and the range of the $q(0)$ sawtooth oscillations are significantly reduced with respect to the reference case. At a time of maximum internal kink amplitude, the radial profile of the $1/1$ $B_r$ harmonic is characterized by the usual peak in the core. In this case, the $B_r$ profile is not zero outside the peak, but increases towards the edge to match the amplitude of the applied MP. The radial profile of the vacuum $1/1$ $B_r$ harmonic is also shown. The vacuum harmonic is larger than the plasma harmonic at the edge. This means that the plasma is partially screening the MP, even in the absence of a mean plasma flow. It is also interesting to note that the external perturbation clamps the phase of the internal kink. In fact, the core $1/1$ $B_r$ amplitude reverses after each crash as in the reference case, but now it reverses again shortly after and always grows positive. The edge $m=1$ perturbation exerts some effect on other harmonics too. In particular, the $2/1$ $B_r$ harmonic is much larger than in the reference case, and it is finite not only during the sawtooth crashes but throughout the entire simulation. This is probably due to both a geometric coupling effect and to a change in the linear stability of the $2/1$ tearing mode, which appears to be close to its stability threshold in this equilibrium configuration. The corresponding Poincar\'e plot is characterized by quite large secondary islands at major resonant surfaces, like in particular the $2/1$ magnetic island. In addition, magnetic chaos appears in the edge region as a result of the overlapping of edge islands. The last closed magnetic surface is located between the island chains at $q=3$ and $q=4$ resonances. A complete mitigation of the sawtoothing activity could be achieved by further increasing the MP amplitude as in the circular tokamak case \cite{bonfiglio2013eps}. However, in this D-shaped case the resulting level of chaos would be too high to be experimentally relevant.


A similar sawtooth mitigation effect is obtained with a $2/1$ MP of the same amplitude, as shown in Fig. \ref{fig:6}. In this case, the radial $B_r$ profile of the perturbed $2/1$ harmonic is close to the vacuum solution near the wall, and it is largely amplified inside $r/a=0.5$ by the clearly unstable $2/1$ tearing mode. The sawtooth mitigation effect is not provided by the direct action on the internal kink mode as in the previous case, but it comes from the action on the $2/1$ tearing mode mediated by the toroidal coupling. In the circular tokamak study with $q_a\simeq 2$ \cite{bonfiglio2013eps}, the $1/1$ MP was somehow more effective than the $2/1$ for the sawtooth mitigation. The fact that in the present study the two perturbations are almost equivalent is reasonable since here the edge $q$ is larger and the $q=1$ resonance is more internal, so that the direct $1/1$ MP is less effective. As far as magnetic topology is concerned, the edge stochastization is stronger in the $2/1$ perturbation case. In fact, the last closed magnetic surface is now in between the island chains at the $q=5/2$ and $q=3$ resonances.


A completely different behaviour is observed in the case of $3/1$ MP, as reported in Fig. \ref{fig:7}. In fact, the average period and the range of the $q(0)$ sawtooth oscillations are larger than in the reference case. The cause of this effect is not clear at the moment. Indeed, the plasma amplification of the perturbed $3/1$ $B_r$ harmonic is relatively small, and the $B_r$ profiles inside $r/a=0.5$ are overall similar to the case with $1/1$ MP. The level of edge chaos is also similar to the $1/1$ perturbation case. Finally, the case with $4/1$ MP is shown in Fig. \ref{fig:8}. The effect of such perturbation on the sawtooth oscillations turns out to be neglibigle in this case. In fact, the MHD dynamics is quite similar to the reference case, including the spontaneous $\pi$ phase flips after each crash. The radial $B_r$ profile of the $4,1$ harmonic is rather close to the vacuum profile, meaning that the $4/1$ mode is stable. The edge stochastization effect is also rather limited. In fact, a relatively narrow chaotic layer is oberved outside the island chain at the $q=5$ resonance.

\section{Final remarks and perspectives}
\label{sec:conclusion}


In this paper, fully nonlinear 3D MHD simulations of shaped tokamak plasmas performed with the \textsc{pixie3d} code in toroidal geometry have been reported. Plasma shaping is obtained by the application of $n=0$ perturbations of the circular magnetic boundary. If the amplitude of the MP is large enough, shaped configurations with X-points can be obtained, of interest for diverted tokamak studies. This approach makes possible to model the magnetic separatrix and even the outside scrape-off layer with more flexibility than MHD codes based on flux coordinates. For instance, an unrealistic truncation of the poloidal flux close to the separatrix is required in the \textsc{mars} code \cite{liu2012ppcf,haskey2015ppcf}. A diverted, D-shaped equilibrium configuration has been considered, and the effect of $n=1$ MPs has been studied. Numerical results confirm the sawtooth mitigation effect with $1/1$ and $2/1$ MPs found in previous circular tokamak simulations. A similar sawtooth mitigation effect with $n=1$ MPs has been also observed in both RFX-mod \cite{bonfiglio2013eps} and DIII-D tokamak plasmas \cite{martin2014iaea}. Such investigations have been triggered by analogous experimental and numerical findings in the RFP configuration.


Although tokamak simulations in this report are performed in the zero-$\beta$ regime and without plasma rotation, these two approximations are going to be removed in the near future. Plasma rotation is known to significantly affect the plasma response to resonant MPs, in particular by suppressing the magnetic islands that would form at resonant surfaces \cite{fitzpatrick1991pofb}. The inclusion of a source in the momentum balance equation has been recently implemented in \textsc{pixie3d} and the effect of the resulting mean toroidal flow is going to be investigated. Finite $\beta$ and the heat transport equation are also going to be switched on, which are expected to affect the plasma equilibrium and the plasma response to MPs, and to possibly destabilize pressure-driven modes. Two-fluid \cite{ferraro2012pop} and kinetic \cite{wang2012pop} effects on MHD modes are also increasingly important when approaching realistic low-collisionality conditions. The coupling of \textsc{pixie3d} with a kinetic module for fast particles, as in the hybrid \textsc{hmgc} code \cite{briguglio1995pop}, will be considered in the future. Overall, the above numerical advancements are expected to significantly improve the predictive capability of the numerical tool. Not only tokamak modelling, but also RFP and stellarator studies \cite{bonfiglio2015ppcf} are going to greatly benefit from such advancements.

\section*{Acknowledgements}

One of the authors (D. B.) would like to thank P. Piovesan for useful discussions. This work was carried out using the HELIOS supercomputer system at Computational Simulation Centre of International Fusion Energy Research Centre (IFERC-CSC), Aomori, Japan, under the Broader Approach collaboration between Euratom and Japan, implemented by Fusion for Energy and JAEA. This project has received funding from the European Union's Horizon 2020 research and innovation programme under grant agreement number 633053. The views and opinions expressed herein do not necessarily reflect those of the European Commission.

\bibliographystyle{iaea}
\bibliography{paper}

\begin{thebibliography}{10}

\bibitem{lorenzini2009np}
LORENZINI, R. et~al.,
\newblock Nature Phys. {\bf 5} (2009) 570.

\bibitem{weller1987prl}
WELLER, A. et~al.,
\newblock Phys. Rev. Lett. {\bf 59} (1987) 2303.

\bibitem{graves2013ppcf}
GRAVES, J.~P. et~al.,
\newblock Plasma Phys. Control. Fusion {\bf 55} (2013) 014005.

\bibitem{chu2010ppcf}
CHU, M.~S. et~al.,
\newblock Plasma Phys. Control. Fusion {\bf 52} (2010) 123001.

\bibitem{evans2004prl}
EVANS, T.~E. et~al.,
\newblock Phys. Rev. Lett. {\bf 92} (2004) 235003.

\bibitem{garofalo2008prl}
GAROFALO, A.~M. et~al.,
\newblock Phys. Rev. Lett. {\bf 101} (2008) 195005.

\bibitem{bortolon2013prl}
BORTOLON, A. et~al.,
\newblock Phys. Rev. Lett. {\bf 110} (2013) 265008.

\bibitem{sonato2003fed}
SONATO, P. et~al.,
\newblock Fus. Eng. Des. {\bf 66} (2003) 161.

\bibitem{piovesan2011ppcf}
PIOVESAN, P. et~al.,
\newblock Plasma Phys. Control. Fusion {\bf 53} (2011) 084005.

\bibitem{piovesan2014nf}
PIOVESAN, P. et~al.,
\newblock Nucl. Fusion {\bf 54} (2014) 064006.

\bibitem{cappello2012iaea}
CAPPELLO, S. et~al.,
\newblock in {\em Proceedings of the 24th IAEA Fusion Energy Conference, San
  Diego, USA}, 2012.

\bibitem{martin2013nf}
MARTIN, P. et~al.,
\newblock Nucl. Fusion {\bf 53} (2013) 104018.

\bibitem{piovesan2013pop}
PIOVESAN, P. et~al.,
\newblock Phys. Plasmas {\bf 20} (2013) 056112.

\bibitem{bonfiglio2013eps}
BONFIGLIO, D. et~al.,
\newblock in {\em Proceedings of the 40th European Physical Society Conference
  on Controlled Fusion and Plasma Physics, Espoo, Finland}, 2012.

\bibitem{martin2014iaea}
MARTIN, P. et~al.,
\newblock in {\em Proceedings of the 25th IAEA Fusion Energy Conference, Saint
  Petersburg, Russia}, 2014.

\bibitem{escande2000ppcf}
ESCANDE, D.~F. et~al.,
\newblock Plasma Phys. Control. Fusion {\bf 42} (2000) B243.

\bibitem{cappello1992pof}
CAPPELLO, S. et~al.,
\newblock Phys. Fluids B {\bf 4} (1992) 611.

\bibitem{cappello2000prl}
CAPPELLO, S. et~al.,
\newblock Phys. Rev. Lett. {\bf 85} (2000) 3838.

\bibitem{cappello2004ppcf}
CAPPELLO, S.,
\newblock Plasma Phys. Control. Fusion {\bf 46} (2004) B313.

\bibitem{bonfiglio2011nf}
BONFIGLIO, D. et~al.,
\newblock Nucl. Fusion {\bf 51} (2011) 063016.

\bibitem{veranda2013ppcf}
VERANDA, M. et~al.,
\newblock Plasma Phys. Control. Fusion {\bf 55} (2013) 074015.

\bibitem{bonfiglio2013prl}
BONFIGLIO, D. et~al.,
\newblock Phys. Rev. Lett. {\bf 111} (2013) 085002.

\bibitem{bonfiglio2015ppcf}
BONFIGLIO, D. et~al.,
\newblock Plasma Phys. Control. Fusion {\bf 57} (2015) 044001.

\bibitem{chacon2004cpc}
CHAC\'ON, L.,
\newblock Comp. Phys. Comm. {\bf 167} (2004) 143.

\bibitem{chacon2008pop}
CHAC\'ON, L.,
\newblock Phys. Plasmas {\bf 15} (2008) 056103.

\bibitem{cappello1996nf}
CAPPELLO, S. et~al.,
\newblock Nucl. Fusion {\bf 36} (1996) 571.

\bibitem{bonfiglio2010pop}
BONFIGLIO, D. et~al.,
\newblock Phys. Plasmas {\bf 17} (2010) 082501.

\bibitem{delzanno2008pop}
DELZANNO, G.~L. et~al.,
\newblock Phys. Plasmas {\bf 15} (2008) 122102.

\bibitem{furth1973pof}
FURTH, H.~P. et~al.,
\newblock Phys. Fluids {\bf 16} (1973) 1054.

\bibitem{shan1993ppcf}
SHAN, X. et~al.,
\newblock Plasma Phys. Control. Fusion {\bf 35} (1993) 619.

\bibitem{wesson1978nf}
WESSON, J.~A.,
\newblock Nucl. Fusion {\bf 18} (1978) 87.

\bibitem{finn2005pop}
FINN, J.~M. et~al.,
\newblock Phys. Plasmas {\bf 12} (2005) 054503.

\bibitem{liu2012ppcf}
LIU, Y.~Q. et~al.,
\newblock Plasma Phys. Control. Fusion {\bf 54} (2012) 124013.

\bibitem{haskey2015ppcf}
HASKEY, S.~R. et~al.,
\newblock Plasma Phys. Control. Fusion {\bf 57} (2015) 025015.

\bibitem{fitzpatrick1991pofb}
FITZPATRICK, R. et~al.,
\newblock Phys. Fluids B {\bf 3} (1991) 644.

\bibitem{ferraro2012pop}
FERRARO, N.~M.,
\newblock Phys. Plasmas {\bf 19} (2012) 056105.

\bibitem{wang2012pop}
WANG, Z.~R. et~al.,
\newblock Phys. Plasmas {\bf 19} (2012) 072518.

\bibitem{briguglio1995pop}
BRIGUGLIO, S. et~al.,
\newblock Phys. Plasmas {\bf 2} (1995) 3711.

\end{thebibliography}

\end{document}